\documentclass{ws-procs9x6}

\begin{document}


\newcommand{\nc}{\newcommand}
\newcommand{\rc}{\renewcommand}
\nc{\ba}{\begin{eqnarray}}
\nc{\ea}{\end{eqnarray}}
\nc{\cal}{\mathcal}
\nc{\eq}[1]{Eq.~(\ref{#1})}
\nc{\eqs}[2]{Eqs.~(\ref{#1}--\ref{#2})}
\nc{\se}[1]{Sec.~\ref{#1}}
\nc{\re}[1]{Ref.~\cite{#1}}
\nc{\e}{\epsilon}
\rc{\a}{\alpha}
\nc{\ta}{{\tilde\alpha}}
\nc{\tb}{{\tilde\beta}}
\nc{\fr}{\frac}
\def\({\left(}
\def\){\right)}
\nc{\lk}{\left[}
\nc{\rk}{\right]}
\nc{\lb}{\left\{}
\nc{\rb}{\right\}}
\nc{\ld}{\left.}
\nc{\rd}{\right.}
\nc{\hP}{{\hat\Pi}}
\nc{\hg}{{\hat g}}
\nc{\hm}{{\hat m}}
\nc{\hb}{{\hat\beta}}
\nc{\hl}{{\hat\lambda}}
\nc{\hn}{{\hat n}}
\nc{\nn}{\nonumber\\}
\nc{\order}{{\cal O}}
\nc{\sumint}{\sum\!\!\!\!\!\!\int}
\nc{\msbar}{$\overline{\rm MS}$}


\title{\vspace{-38pt}Evading the Infrared Problem of Thermal QCD\footnote{
\uppercase{P}artially supported by the \uppercase{DOE},  
\uppercase{C}ooperative \uppercase{A}greement 
no.~\uppercase{DF}-\uppercase{FC}02-94\uppercase{ER}40818.}}

\author{Y. SCHR\"ODER}

\address{Center for Theoretical Physics, MIT, Cambridge, MA 02139, USA\\
and\\
Faculty of Physics, University of Bielefeld, D-33501 Bielefeld, Germany\\[10pt]
\mbox{}\hfill MIT-CTP-3543, BI-TP 2004/27}

\maketitle

\abstracts{
Due to asymptotic freedom, QCD is guaranteed to be accessible to perturbative
methods at asymptotically high temperatures. However, in 1979 Linde has
pointed out the existence of an "infrared wall", beyond which an infinite
number of Feynman diagrams contribute. Following a proposal by Braaten and
Nieto, it is shown explicitly how the limits to computability that this
infrared problem poses can be overcome in the framework of dimensionally
reduced effective theories.}


\addtolength{\textheight}{-12pt}

\section{Introduction}

The theory of strong interactions, Quantum Chromodynamics (QCD), 
is guaranteed to be accessible to perturbative methods once
one of its parameters, the temperature $T$, is increased to
asymptotically high values. 
This statement relies solely on the well-known property of 
asymptotic freedom\cite{Nobel04}. 

In practice, however, calculations of corrections to the behavior 
of an ideal gas of quarks and gluons, 
the limit that is formally realized at infinite $T$,
are obstructed by infrared divergences\cite{Linde}:
For every observable 
one sets out to compute, there exists an order of the perturbative
expansion to which an infinite number of Feynman diagrams contribute
(``infrared wall'').

No method is known how to re-sum these infinite classes of diagrams,
a fact that seriously obstructs progress in the field of thermal
QCD, a field that presently receives attention particularly
due to its relevance to the ongoing program of heavy-ion collisions
at RHIC, where one of the 
main focuses is to explore the phase diagram of QCD. 

For the QCD pressure -- as an example of a static thermodynamic 
observable -- it has been shown explicitly how the limits to 
computability that the infrared problem poses can be overcome 
in the framework of dimensionally reduced effective 
theories\cite{BN,Howto,g6lng}. The key idea is to 
measure the effect of ultra-soft (``magnetic'') gluons
by lattice Monte--Carlo simulations in 3--dimensional (3d) 
pure gauge theory (MQCD), and to match this theory up 
to full thermal QCD in perturbation theory, 
via 3d gauge + adjoint Higgs theory (EQCD).

The expansion of the QCD pressure in the effective theory
framework, up to the order where infrared contributions are relevant, 
is now known analytically\cite{g6lng}. 
The $\order(g^6)$ coefficient is non--perturbative, 
but {\em computable}.
All other effects ($N_f$- or $\mu_f$-dependence; 
orders $g^7$, ...) are perturbative.


\section{Status of the QCD Pressure}

Below, we specify the contributions to the pressure 
$p_{\rm QCD}=p_G+p_M+p_E$ from
each physical scale individually, for the slightly more general case of 
gauge group SU($N_c$) and $N_f$ quark flavors, mainly following 
\re{g6lng}.
We will work at zero quark masses $m_{q_i}=0$ and vanishing 
chemical potential $\mu_f=0$,
and display all dependence on the \msbar~scale by 
$L\equiv\ln\fr{\bar\mu}{4\pi T}$.

$\bullet$ Contributions from the ultra-soft scale $g^2 T$, i.e. from MQCD:
\ba
\fr{p_G(T)}{\mu^{-2\e}} &=&
 d_A 16\pi^2 T^4 \hg_M^6 \lk \a_G\(\fr1\e+8L-8\ln(8\pi\hg_M^2)\) +\beta_G
 +\order(\e) \rk , 
\ea
where $d_A=N_c^2-1$, 
$\a_G=\fr{43}{96}-\fr{157}{6144}\pi^2$
is a perturbative 4-loop coefficient, 
and $\beta_G$ is non-perturbative, requiring two ingredients: 
a lattice-measurement in 
MQCD, and a perturbative computation which allows
to match between the lattice and the continuum regularization schemes.
For the latter, a 4-loop lattice-regularized computation is needed,
which could possibly be accomplished with methods used 
in Refs.~\cite{nspt,pana}.
The matching condition reads
\ba
\label{eq:gM}
 \hg_M^2\equiv\fr{N_c g_M^2}{16\pi^2 T} = 
 \hg_E^2 +\order(\hg_E^4 \hm_E^{-1}) \;.
\ea

$\bullet$ Contributions from the soft scale $g T$, i.e. from EQCD:
\ba
\label{eq:pM}
\fr{p_M(T)}{\mu^{-2\e}} &=&
 d_A 16\pi^2 T^4 \lb 
 \hm_E^3\lk \fr13 +\order(\e) \rk
\rd\nn&&{}\ld
 +\hg_E^2\hm_E^2\lk -\fr1{4\e} +\(-L+\fr12\ln\hm_E^2+\ln2-\fr34\) 
 +\order(\e) \rk
\rd\nn&&{}\ld
 +\hg_E^4\hm_E\lk \(-\fr{89}{24}-\fr{\pi^2}6+\fr{11}6\ln2\) +\order(\e) \rk 
\rd\nn&&{}\ld
 +\hg_E^6\lk \a_M\(\fr1\e+8L-4\ln\hm_E^2-8\ln2\) +\beta_M +\order(\e) \rk 
\rd\nn&&{}\ld
 +\hl_E^{(1)}\hm_E^2\lk \fr{\hn-2}4 +\order(\e) \rk 
 +\hl_E^{(2)}\hm_E^2\lk \fr{1-3\hn}4 +\order(\e) \rk 
\rd\nn&&{}\ld
 +\order(\hg_E^8 \hm_E^{-1},\hl_E^2\hm_E) 
\rb ,
\ea
with $\hn\equiv\fr{N_c^2-1}{N_c^2}$, the 4-loop coefficients 
$\a_M = \fr{43}{32}-\fr{491}{6144}\pi^2$,
$\beta_M \approx -1.391512$
and the matching parameters
\ba
 \label{eq:mE}
 \hm_E^2 \equiv 
 \(\fr{m_E}{4\pi T}\)^2 &=& 
 \hg^2 \lk \ta_{\rm E4} +\(2\ta_{\rm E4}L+\ta_{\rm E5}\)\e +\order(\e^2) \rk 
\nn &&
 +\hg^4\lk \( 2\hb_0\ta_{\rm E4} L+\ta_{\rm E6} \) +\bar\beta_{\rm E2}\e 
 +\order(\e^2) \rk
 +\order(\hg^6) \,, \\
\hg_E^2 \equiv
 \fr{N_c g_E^2}{16\pi^2 T} &=& 
 \hg^2 
 + \hg^4\lk \( 2\hb_0 L +\ta_{\rm E7} \) +\bar\beta_{\rm E3}\e 
 +\order(\e^2) \rk 
 +\order(\hg^6) \,, \\
\hl_E^{(1)} \equiv 
 \fr{N_c^2 \lambda_E^{(1)}}{16\pi^2 T} &=& 
 \hg^4\lk 4 +\order(\e) \rk +\order(\hg^6) \;, \\
\hl_E^{(2)} \equiv
 \fr{N_c \lambda_E^{(2)}}{16\pi^2 T} &=& 
 \hg^4\lk \fr43(1-z) +\order(\e) \rk +\order(\hg^6) \;,
\ea
where, for brevity, $z\equiv N_f/N_c$ and
\ba
\label{eq:aE4}
\ta_{\rm E4} = \fr{2+z}6 \;, \quad 
\ta_{\rm E6} = \fr13\ta_{\rm E4}(6\hb_0\gamma+5+2z-8z\ln2) 
 -\fr{z}2\hn \;,\\
\label{eq:aE6}
\ta_{\rm E5} = 2\ta_{\rm E4}Z_1 +\fr{z}6(1-2\ln2) \;, \quad
\ta_{\rm E7} = 2\hb_0\gamma +\fr13 -\fr83z\ln2 \;,
\ea
and $\bar\beta_{\rm E2}$ (see \se{se:bE2}) and $\bar\beta_{\rm E3}$ 
remain to be computed
by perturbatively matching suitable correlators computed in 
thermal QCD and in EQCD. 

$\bullet$ Contributions from the hard scale $2\pi T$, i.e. from thermal QCD:
\ba
\fr{p_E(T)}{\mu^{-2\e}} &=&
 d_A 16\pi^2 T^4 \fr1{16}\fr1{45} \lb 
 \lk 1 +\fr74\fr{z}{\hn} \rk 
 +\hg^2\lk \ta_{\rm E2} +\order(\e) \rk 
\rd\nn&&{}\ld
 +\hg^4\lk \ta_{\rm E4}\fr{180}{\e} 
 +(180\cdot 6\ta_{\rm E4}+2\hb_0\ta_{\rm E2} )L +\ta_{\rm E3} +\order(\e) \rk 
\rd\nn&&{}\ld
 +\hg^6\lk 
 \fr{\tb_{\rm E1}^{\rm (div)}}{\e} 
 +\tb_{\rm E1}^{(L^2)}L^2
 +\tb_{\rm E1}^{(L)}L
 +\tb_{\rm E1} 
 +\order(\e) \rk
 +\order(\hg^8) \rb ,\quad
\ea
with $\ta_{\rm E2} = -\fr54(4+5z)$ and, writing 
$Z_1\equiv\fr{\zeta^\prime(-1)}{\zeta(-1)}$ and
$Z_3\equiv\fr{\zeta^\prime(-3)}{\zeta(-3)}$,
\ba
\ta_{\rm E3} &=& 
 180(\ta_{\rm E4})^2\gamma
 +5\lk \(\fr{116}5+\fr{220}3Z_1-\fr{38}3Z_3\) 
\rd\nn&&{}\ld
 +\fr{z}2\(\fr{1121}{60}-\fr{157}5\ln2+\fr{146}3Z_1-\fr13Z_3\) 
\rd\nn&&{}\ld
 +\fr{z^2}4\(\fr13-\fr{88}5\ln2+\fr{16}3Z_1-\fr83Z_3\) 
 +\fr{z}4\hn\(\fr{105}4-24\ln2\) \rk ,\quad
\ea
and unknown coefficients $\beta_{\rm E1}$, which can be determined
e.g. by a 4-loop computation of vacuum diagrams in thermal QCD.
Since $p_{\rm QCD}$ is physical, 
the divergent and scale-dependent parts of $\beta_{\rm E1}$
are related to the other coefficients introduced in the above,
serving as a valuable check on this open computation.
Specifically, from 2-loop running of the 4d gauge coupling
\ba
\hg^2 \equiv \fr{N_c g^2(\bar\mu)}{16\pi^2} &=&
 \hg^2(\bar\mu_0) +\hg^4(\bar\mu_0)(-2\hb_0\ell) 
 +\hg^6(\bar\mu_0)(4\hb_0^2\ell^2-2\hb_1\ell) 
\ea
with beta-function coefficients 
$\hb_0=\fr{11-2z}3$,
$\hb_1=\fr{34}3-\fr{10}3z-z\hn$,
\msbar~scale parameter $\bar\mu^2 = 4\pi e^{-\gamma}\mu^2$
and 
$\ell \equiv \ln\fr{\bar\mu}{\bar\mu_0} = L-\ln\fr{\bar\mu_0}{4\pi T}$,
one can fix
\ba
\tb_{\rm E1}^{\rm (div)} &=& 
 180\lk 4\hb_0\ta_{\rm E4} L +\ta_{\rm E6} +\ta_{\rm E4}\ta_{\rm E7}
 -4(\a_G+\a_M) \rk , \\
\tb_{\rm E1}^{(L^2)} &=& 
 180\lk 20\hb_0\ta_{\rm E4} +\tb_{\rm E2}^{(L^2)} 
 +\ta_{\rm E4}\tb_{\rm E3}^{(L^2)} \rk +4\hb_0^2\ta_{\rm E2} \;, \\
\tb_{\rm E1}^{(L)} &=& 
 180\lk 4\ta_{\rm E6} +6\ta_{\rm E4}\ta_{\rm E7} -2\hb_0\ta_{\rm E5} 
 -32(\a_G+\a_M) +\tb_{\rm E2}^{(L)} +\ta_{\rm E4}\tb_{\rm E3}^{(L)} \rk 
\nn&&{}
 +2\hb_1\ta_{\rm E2} +4\hb_0\ta_{\rm E3} \;. 
\ea


\section{Determination of $\beta_{E2}$}\label{se:bE2}

To determine $\bar\beta_{E2}$ in \eq{eq:mE}, one can e.g. 
match the pole masses of the $A_0$ propagator. 
In thermal QCD, writing $\Pi_{00}^{ab}(k_0=0,\vec k)=\delta^{ab}\Pi(k^2)$,
one needs to solve $k^2+\Pi(k^2)=0$ at $k^2=-m_{\rm pole}^2$.
Inserting a loop expansion for the self-energy $\Pi$ and noting that
the leading-order solution gives a perturbatively small 
$m_{\rm pole}^2\sim g^2$, one can Taylor-expand to get
\ba
\label{eq:mpole}
\hm^2_{\rm pole} &\equiv&
 \(\fr{m_{\rm pole}}{4\pi T}\)^2 =
 \hP_1(0)+\hP_2(0)-\hP_1(0) \Pi_1^\prime(0)+\order(\hg^6) \;.
\ea
The bare gluon self-energies can be deduced from the literature\cite{BN} as
\ba
\hP_1(0) \equiv 
 \fr{\Pi_1(0)}{16\pi^2 T^2} &=& 
 \hg^2_B \lk (d-2)^2 \hat I_b(1) -2z(d-2)\hat I_f(1) \rk ,\\
\Pi_1^\prime(0) &=&
 \hg^2_B \lk \fr16 (-22+7d-d^2) \hat I_b(2) +\fr z3(d-2)\hat I_f(2) \rk ,\\
\hP_2(0) \equiv 
 \fr{\Pi_2(0)}{16\pi^2 T^2} &=& 
 \hg^4_B \lk \( 2z\hat I_f(1)  -(d-2)\hat I_b(1) \) (12-8d+d^2) \hat I_b(2) 
 \rd \nn && {} \ld
 +\hn z(d-4)(d-2) \( \hat I_b(1) - \hat I_f(1) \) \hat I_f(2) 
\rk ,
\ea
where we have used scaled bosonic and fermionic 1-loop tadpole integrals 
$\hat I(x)=16\pi^2(4\pi T)^{2x-4}
T\sum_{n=-\infty}^{\infty}\mu^{2\e}\int\fr{d^{d-1}p}{(2\pi)^{d-1}}
\fr1{(\omega_n^2+\vec p^2)^x}$, 
with $\omega_n=2n\pi T$ for bosons 
and $(2n+1)\pi T$ for fermions,
\ba
\hat I_b(x) &=& \mu^{2\e} \fr{2^{2x-3}}{\sqrt{\pi}}(\sqrt{\pi}T)^{d-4}
 \zeta(1+2x-d)\fr{\Gamma(x+\fr{1-d}2)}{\Gamma(x)} 
\ea
and correspondingly $\hat I_f(x) = (2^{2x+1-d}-1) \hat I_b(x)$. 

In EQCD, the solution of $k^2+m_E^2+\Pi_E(k^2)=0$ at $k^2=-m_{\rm pole}^2$
is simply $m_{\rm pole}^2=m_E^2$, since again treating $m_{\rm pole}^2$ as
perturbatively small and Taylor-expanding, there is no scale left in $\Pi_E$, 
such that it vanishes in \msbar.
Renormalizing \eq{eq:mpole} via 
$\hg^2_B=Z^2_{\hg}\hg^2$
with $Z^2_{\hg}=1-\hg^2 \hb_0 / \e +\order(\hg^4)$
and comparing with \eq{eq:mE},
one reproduces \eqs{eq:aE4}{eq:aE6}
and finally obtains the coefficients in
$\bar\beta_{\rm E2}= 6\hb_0\ta_{\rm E4} L^2 + \tb_{\rm E2}^{(L)} L 
+ \tb_{\rm E2}$, which are
\ba
\tb_{\rm E2}^{(L)} &=& 
 4\hb_0\ta_{\rm E4}(2\gamma+Z_1)
 +\fr19(20+29z+2z^2)
 \nn&&{}
 -2z\(\hn+3\ln2\)
 -\fr43 z^2 \ln2 \;, \\
\tb_{\rm E2} &=& 
 \fr14\hb_0\ta_{\rm E4}\(16\zeta^\prime(1)+\pi^2\)
 +\fr23\ta_{\rm E4}Z_1(6\hb_0\gamma+5+2z-8z\ln2)
 \nn&&{}
 +\fr29\gamma(5+10z-(19+2z)z\ln2)
 +\fr29
 +\fr{z}{18}(7+6\ln2-16\ln^22)
 \nn&&{}
 +\fr{z^2}9(1-2\ln2+4\ln^22)
 -\fr{z}6\hn(3+6\gamma+6Z_1+10\ln2) \;.
\ea


\section{Outlook}

The future should see a completion of the above setup, 
thereby establishing a first
example of successfully computing an observable beyond the infrared wall. 
Once this is achieved, thermal QCD in its high-temperature phase
will again be amenable to perturbative calculations, opening up
numerous opportunities to precisely compute observables that might
become relevant to the RHIC program, to future accelerators, 
and to cosmology.
Indeed, the next term in the series, formally of order $\order(g^7)$, 
requires 
the corrections of order $\order(\hg_E^4/\hm_E)$ to \eq{eq:gM} (known), 
of order $\order(\hg_E^8/\hm_E)$ to \eq{eq:pM} (5-loop vacuum diagrams
in EQCD), 
and of order $\order(\hg^6)$ to \eq{eq:mE} (3-loop 2-point functions
in thermal QCD),
and will then certainly be within reach.

%
%



\begin{thebibliography}{0}

\bibitem{Nobel04}
{\tt http://nobelprize.org/physics/laureates/2004}

\bibitem{Linde}
A.~D.~Linde,
Phys.\ Lett.\ B {\bf 96} (1980) 289;
D.~J.~Gross, R.~D.~Pisarski and L.~G.~Yaffe,
Rev.\ Mod.\ Phys.\  {\bf 53} (1981) 43.

\bibitem{BN} 
E.~Braaten and A.~Nieto,
Phys.\ Rev.\ D {\bf 53} (1996) 3421.

\bibitem{Howto}
K.~Kajantie et. al., 
Phys.\ Rev.\ Lett.\  {\bf 86} (2001) 10.

\bibitem{g6lng}
K.~Kajantie et.al., 
Phys.\ Rev.\ D {\bf 67} (2003) 105008. 

\bibitem{nspt}
F.~Di Renzo, A.~Mantovi, V.~Miccio and Y.~Schroder,
JHEP {\bf 0405} (2004) 006.

\bibitem{pana}
B.~Alles, M.~Campostrini, A.~Feo and H.~Panagopoulos,
Phys.\ Lett.\ B {\bf 324} (1994) 433.

\end{thebibliography}
\end{document}